\documentclass{optica-article}

\journal{opticajournal} 

\articletype{Research Article}

\usepackage{lineno}
\usepackage{xcolor}
\usepackage{float}

\begin{document}

\title{Nonlinear Frequency Translation in Micromachined Rb Vapor Cells}



\author{Heleni Krelman,\authormark{1} Ori Nefesh,\authormark{1} and Liron Stern\authormark{1,*}}

\address{\authormark{1}Institute of Applied Physics, the Center for Nanoscience and Nanotechnology, Hebrew University of Jerusalem, Jerusalem 91904, Israel}

\email{\authormark{*}liron.stern@mail.huji.ac.il } 


\begin{abstract*} 
The exceptional nonlinearity of alkali-metal vapors enables highly efficient nonlinear optical processes even at relatively low optical intensities. However, such processes have traditionally relied on centimeter-scale vapor cells. Here, we utilize a versatile chip-scale Rb vapor platform to generate coherent blue and mid-IR light in continuous-wave mode by means of resonant four-wave mixing. Optimized optical overlap with the atomic medium enables blue light generation of {$\sim$}20 {$\mu$}W over a very short interaction length, while maintaining a directly measured linewidth of {$\sim$}1 MHz, which is presently limited by the measurement apparatus. Comparison with a conventional glassblown vapor cell further shows that the micromachined platform can achieve higher coherent blue-light generation efficiency despite its substantially shorter interaction length. Moreover, an anodically bonded Si window enables to detect coherent mid-IR emission with collected powers of {$\sim$}50 nW. We further characterize the temperature dependence and input-power scaling of the blue emission, confirming efficient nonlinear conversion within these compact vapor cells. This chip-scale platform provides a versatile foundation for a range of nonlinear optical functions, from precise wavelength references and quantum light sources to next‑generation quantum sensors.

\end{abstract*}




\section{Introduction}
Alkali-metal vapors provide a uniquely strong platform for nonlinear optics because the resonant character of atomic transitions gives rise to exceptionally large nonlinear optical responses. This resonance-driven nonlinearity enables a wide range of applications, including quantum information processing \cite{5,6,7}, development of single-photon sources \cite{8,9,34}, electromagnetically induced transparency (EIT) \cite{49}, and frequency conversion via Rydberg states \cite{1,2,3,4}. \par
Specifically, four-wave mixing (FWM) is a nonlinear optical process governed by the medium’s third-order susceptibility $\chi^{(3)}$, in which the interaction of two or three optical fields gives rise to one or two new frequency components, respectively. Alkali vapors, particularly those of rubidium (Rb) and cesium (Cs), possess inherently large third-order nonlinear optical susceptibilities. Under near-resonant conditions, their $\chi^{(3)}$ values can exceed those of conventional nonlinear materials by several orders of magnitude \cite{38,39,40,41}. This large nonlinearity enables highly efficient FWM and related nonlinear processes even at relatively low optical intensities, and makes alkali vapors excellent candidates for next-generation chip-scale light sources.\par
\begin{figure}[h!]
\centering\includegraphics[width= 12.5 cm]{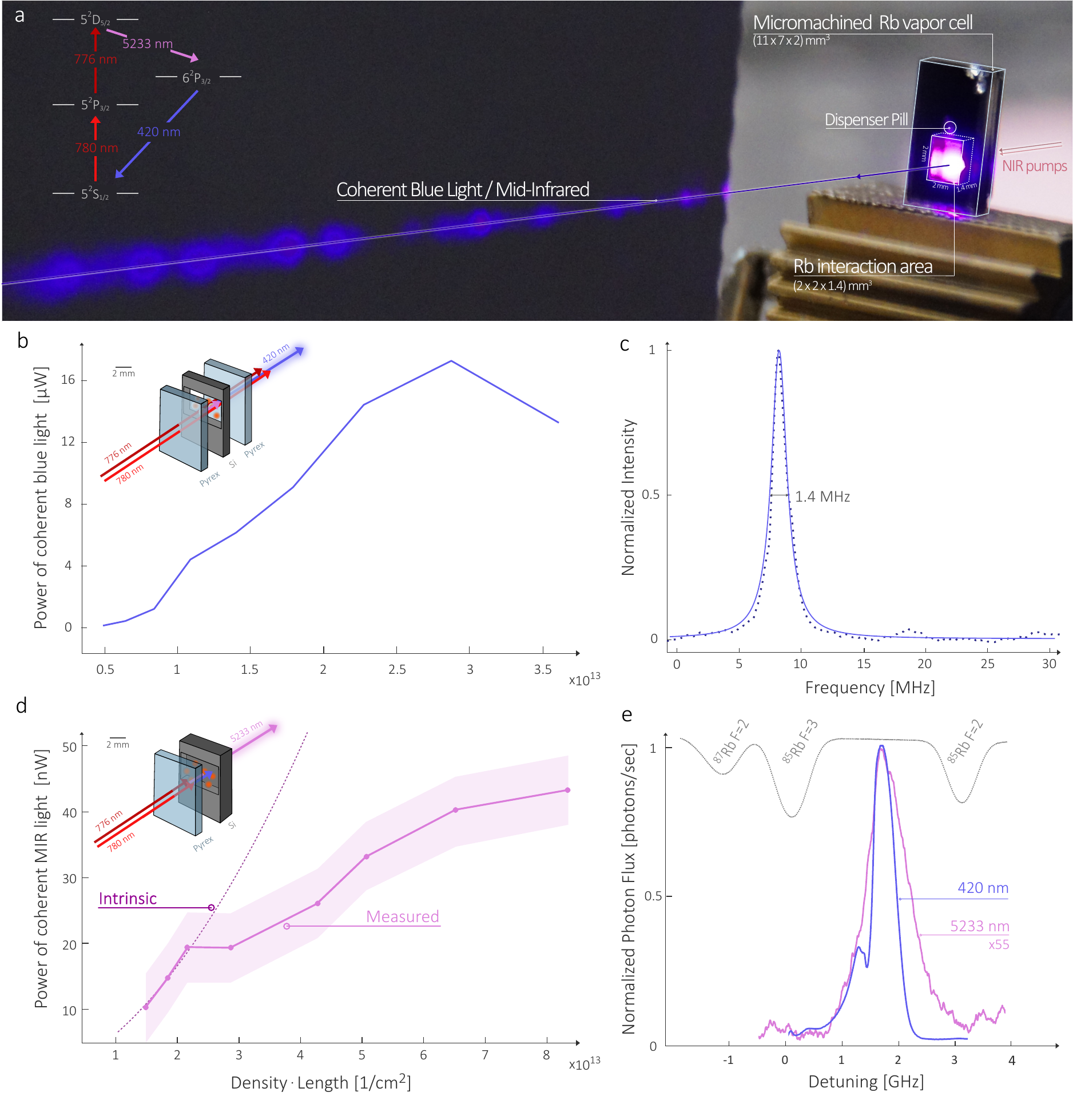}
\caption{\label{fig:fig1} Concept of CW FWM process in chip-scale Rb vapor cell. (a) Photograph of generated blue light through FWM in micromachined Rb vapor cell. For clarity, the captions were added to the photo. By choosing the type of the cell, we can detect either CBL or CMIRL. Inset: Energy levels of \textsuperscript{85}Rb illustrating the transitions involved in generating coherent blue light (CBL) and coherent mid-IR light (CMIRL). The diagram highlights the excitation processes using the 780 nm and 776 nm transitions, leading to production of the 420 nm and 5.2 {$\mu$}m emissions. (b) CBL power versus vapor density-length product. Inset: schematic view of the chip-scale Rb cell with Pyrex backside, which absorbs 5.2 {$\mu$}m emission and transmits the other three wavelengths involved in the process. (c) Spectral linewidth of the 420 nm emission. The dashed line represents the measured spectrum, while the solid line shows a Lorentzian fit, yielding a linewidth of 1.4$\pm$0.2  MHz. (d) CMIRL power versus vapor density-length product. The measured data are represented by circle marks, while the dashed line indicates the intrinsic behavior. Inset: schematic view of the chip-scale Rb cell with Si backside, which transmits 5.2 {$\mu$}m emission only. (e) Photon flux as a function of the 780 nm laser detuning for the measured CBL (blue curve) and CMIRL (purple curve), together with a Rb reference spectrum (gray curve). The CMIRL flux is scaled by a factor of 55 relative to the CBL to account for losses, as discussed in the text.}
\end{figure}
 In this work, we investigate CW FWM in micromachined chip-scale Rb vapor cells, focusing specifically on a diamond four-level configuration that simultaneously generates coherent blue light (CBL) and coherent mid-IR light (CMIRL), as illustrated in the inset of Fig. \ref{fig:fig1}(a). We employed two cell designs with different backside materials: a Pyrex backside that absorbs CMIRL while transmitting CBL, and a Si backside that transmits CMIRL only (the insets of Fig. \ref{fig:fig1}(b) and Fig. \ref{fig:fig1}(d)). Using these cells, we detect both CBL (blue curve in Fig. \ref{fig:fig1}(e)) and CMIRL (purple curve in Fig. \ref{fig:fig1}(e)), reaching maximal powers of 17 {$\mu$}W and 50 nW, respectively. We then systematically compare the scaling of the FWM process between centimeter-scale and chip-scale Rb vapor cells to quantify the impact of miniaturization. We further study the conversion efficiency as a function of temperature (Fig. \ref{fig:fig1}(b) and Fig. \ref{fig:fig1}(d)) and optical intensity in the chip-scale geometry, and measure a linewidth of 1.4$\pm$0.2 MHz for the coherent 420 nm emission (Fig. \ref{fig:fig1}(c)). These results demonstrate a viable route to efficient chip-scale nonlinear light conversion based on atomic vapor systems, with direct relevance to precision spectroscopy and metrology.
 
Traditionally, continuous-wave (CW) FWM has been implemented in centimeter-scale vapor cells, where long interaction lengths enable efficient nonlinear conversion \cite{17,29,30,31,32}. More recently, millimeter-scale vapor cells have demonstrated nonlinear optical functionality in compact atomic systems, primarily for photon-pair generation \cite{33}. Despite this progress, achieving robust CW FWM in chip-scale atomic vapor platforms remains an open challenge. Addressing this challenge is particularly compelling given the broad interest in the generated wavelengths, which are relevant for precision spectroscopy \cite{25}, Rydberg excitation \cite{27,28,50}, thermal imaging \cite{26}, and a range of biomedical applications \cite{21,22,23,24}.
\par

Chip-scale microfabricated alkali vapor cells, fabricated using Si-based processes rather than traditional glass blowing, have emerged as a promising platform for a wide range of applications, including atomic magnetometry \cite{15,16} and electrometry \cite{44, 48}, compact gradiometers \cite{36}, quantum memory storage \cite{46}, offset wavelength references \cite{45}, and atomic clocks \cite{13,14}. These devices are both mass-producible and cost-effective, and offer exceptional miniaturization, high precision, stability, and a clear path for integration with photonic circuits \cite{10,11,12}.\par

\section{Concept and Experimental System}
We briefly outline the FWM mechanism and the specific excitation pathways employed in this study. The excitation of atoms from the ground state 5\textsuperscript{2}S\textsubscript{1/2} to the 5\textsuperscript{2}D\textsubscript{5/2} begins with a ladder excitation utilizing the D\textsubscript{2} transition (780 nm) between the 5\textsuperscript{2}S\textsubscript{1/2} and 5\textsuperscript{2}P\textsubscript{3/2} states and the subsequent 5\textsuperscript{2}P\textsubscript{3/2} to 5\textsuperscript{2}D\textsubscript{5/2} with a wavelength of 775.98 nm, as illustrated in the inset of Fig. \ref{fig:fig1}(a) with relevant levels for \textsuperscript{85}Rb. Due to the longer lifetime of the 5\textsuperscript{2}D\textsubscript{5/2} state relative to the 6\textsuperscript{2}P\textsubscript{3/2} state, the 5\textsuperscript{2}D\textsubscript{5/2} level accumulates a significant population, thus establishing a population inversion that facilitates the generation of mid-IR photons at 5.23 {$\mu$}m via amplified spontaneous emission (ASE). These photons serve as a seed for the FWM process, ultimately resulting in the generation of coherent light both at the visible (420 nm) and mid-IR (5.2 {$\mu$}m) wavelength regimes under phase-matching conditions.\par
As a general guiding principle, we aim to maintain a constant effective FWM gain under miniaturization, starting by controlling the product of the vapor density and the interaction length. The effective gain plays a significant role in the FWM conversion efficiency, defined as the fraction of the injected pump power that is converted into the generated field, that is, the generated CBL/CMIRL power divided by the total input pump power. The FWM conversion efficiency is governed by the phase-matching condition, which by itself depends on the detuning of the injected light sources, the input laser powers and the product of the atomic density \textit{N} and the interaction length \textit{L};  specifically, the efficiency scales as {$sinh^2(\gamma \cdot L)$} where the gain coefficient {$\gamma$} is proportional to the atomic density \cite{35}. This expression corresponds to an approximate analytical solution in the non-depleted pump regime, assuming that both the signal and idler waves are launched at {$z = 0$}. \par
Accordingly, to maintain a constant effective gain (i.e., {$N \cdot L$}), a mm-scale Rb vapor cell must operate at a higher temperature to increase \textit{N}, compared to traditional centimeter-scale glassblown vapor cells \cite{31}.\par

\begin{figure}[h!]
\centering\includegraphics[width=\textwidth]{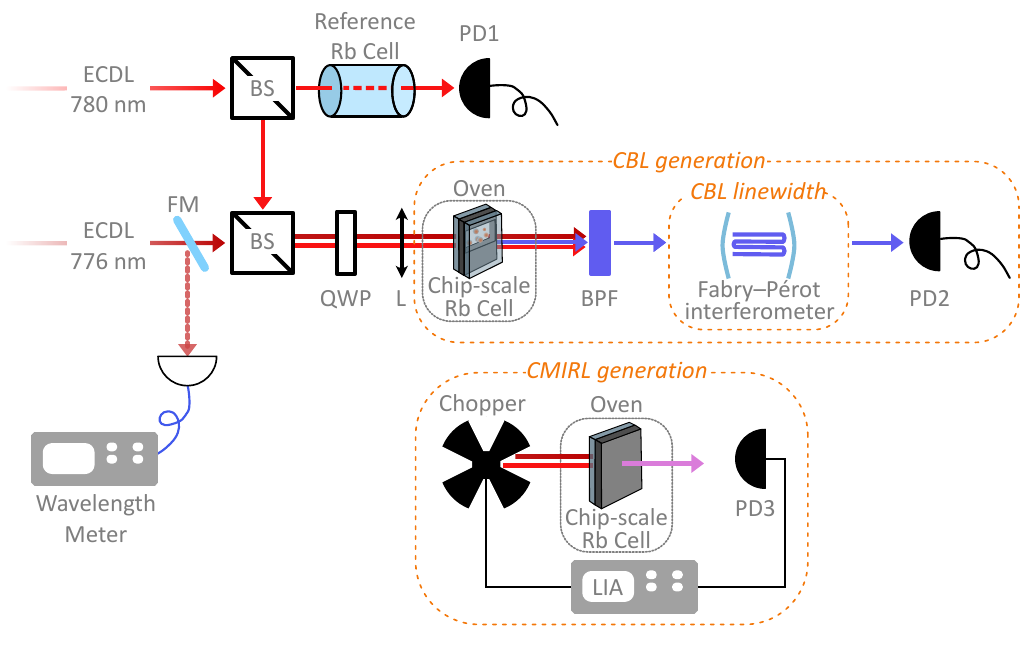}
\caption{\label{fig:fig2}
Schematic of the experimental setup. The 780 nm ECDL beam is split by a beam splitter (BS): one branch is directed to a reference Rb cell and detected by PD1, while the other is combined with the 776 nm ECDL beam. Both beams are circularly polarized and focused by a lens (L) into the Rb cell, where nonlinear processes generate emission at 420 nm and 5.2 {$\mu$}m. The 420 nm light is isolated using a band-pass filter (BPF) and detected by PD2. Detection of the 5.2 {$\mu$}m emission is achieved by optical chopping and lock-in demodulation of the PD3 signal. For CBL linewidth measurements, an external Fabry-Perot cavity is placed after the cell, and the transmitted spectrum is detected by PD2.}
\end{figure}
Fig. \ref{fig:fig2} illustrates our optical arrangement for generating coherent blue and mid-IR emissions via the FWM process in a chip-scale Rb vapor cell. For 420 nm generation, a micromachined Si wafer is anodically bonded to Pyrex on both sides under vacuum, with a Rb molybdate pill incorporated inside. The Si wafer, which also defines the nonlinear interaction length, is 1.4 mm thick. In contrast, for the 5.2 {$\mu$}m cell, the Si wafer is first etched using deep reactive ion etching (DRIE) to create a cavity for the pill and Rb vapor. The etching process defines a blind cavity with an effective interaction length of approximately 1.4 mm. After anodic bonding to Pyrex on one side, the structure is diced into chips with outer dimensions of 11 × 7 × 2 mm\textsuperscript{3} and an internal cavity of 2 × 2 × 1.4 mm\textsuperscript{3}. Although this limitation is not fundamental to the process, the present fabrication approach produces a relatively rough backside, which limits the collection efficiency, as discussed below.\par
The two laser sources are external-cavity diode lasers (ECDLs) operating at 780 nm and 776 nm, each delivering approximately 30 mW of power to the cell. The beams are combined using a beam splitter and circularly polarized to optimize the nonlinear interaction \cite{31}. They are then focused by a lens (L) into the chip-scale Rb vapor cell, where the nonlinear process generates coherent blue light at 420 nm and mid-IR emission at 5.2 {$\mu$}m. In the CBL generation setup, the 420 nm light is isolated by a band-pass filter (BPF) and detected by a Si photodiode (PD2). As mentioned before, in the CMIRL generation setup, a Si backside chip-scale Rb cell is used, and the 5.2 {$\mu$}m light is isolated and detected by an InAsSb photodiode (PD3). Given the relatively high noise floor of available mid-IR detectors, we modulate the signal using an optical chopper and employ a lock-in amplifier (LIA) to improve the signal-to-noise ratio (SNR). Measurements were performed with two lens configurations: one with a 150-mm focal length (FL), resulting in a waist diameter of 360 {$\mu$}m, and another with a 15-mm focal length, resulting in a waist diameter of 120 {$\mu$}m. In the 150-mm configuration, a 7-cm traditional glassblown Rb cell was also used for comparative measurements with the chip-scale cell. Additionally, in our measurements, we track the frequencies that yield maximal CBL emission. To do so, we measure the detuning of the 780 nm laser using a reference Rb cell, while the detuning of the 776 nm laser is monitored with a wavelength meter.\par
To measure the spectral linewidth of the CBL, we employed an external Fabry-Perot cavity with a free spectral range (FSR) of 1.5 GHz and a resolution of approximately 1 MHz. The cavity was positioned directly after the Pyrex-backside vapor cell. The transmission through the cavity was detected using PD2. Mode matching between the spatial mode of the CBL and that of the cavity was achieved by inserting an additional lens with a focal length of 50 mm before the Fabry-Perot cavity.

\section{Results}
The cross section of the laser, the temperature and length of the cell determine the number of atoms participating in the nonlinear interaction, while the intensity of the driven laser plays a critical role in setting the strength of the atomic response. Specifically, higher laser intensity increases the Rabi frequency, which governs how strongly the atoms are driven.\par 
In this work, the micromachined chip-scale vapor cell confines Rb atoms in an extremely small interaction volume, yet the strong atomic nonlinearity enables efficient FWM even in this compact geometry. By systematically varying these parameters, we can investigate different regimes of FWM and compare the resulting efficiencies and spectral characteristics under various conditions.
\subsection{Coherent Blue Emission}
\begin{figure}[h!]
\centering\includegraphics[width=\textwidth]{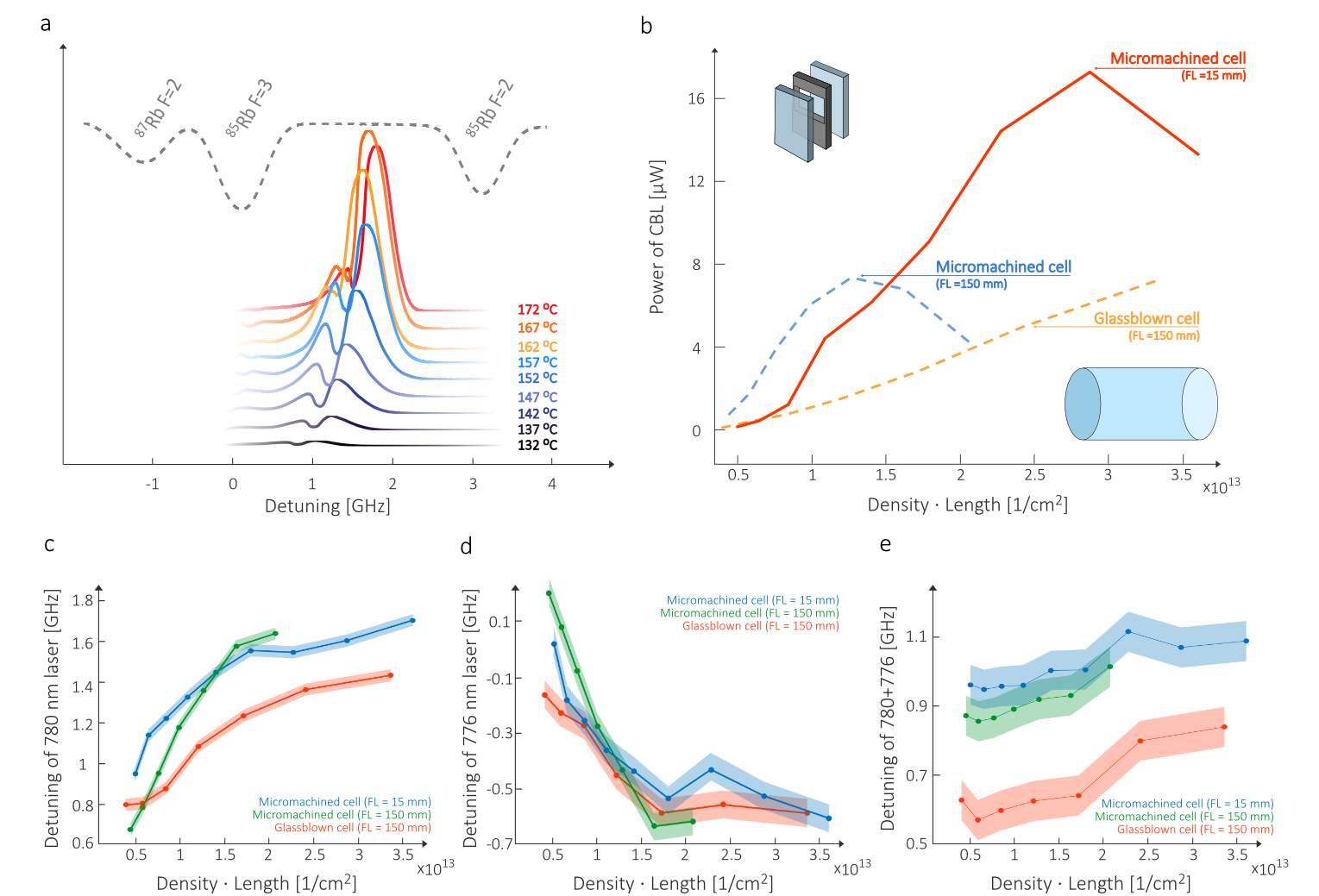}
\caption{\label{fig:fig3} Temperature dependence of CBL. (a) CBL spectra measured in the micromachined cell while scanning the 780 nm laser at various temperatures (vertically offset for clarity), together with a reference Rb absorption signal shown in gray. (b) CBL power versus vapor density-length product for micromachined cells with 15 mm (red) and 150 mm (dashed blue) lenses, and for a glassblown cell with a 150 mm lens (dashed orange). (c-e) Optimal detuning of the (c) 780 nm laser, (d) the 776 nm laser, and (e) their sum versus density-length product for three configurations: a micromachined cell (FL = 15 mm; blue curve), a micromachined cell (FL = 150 mm; green curve), and a glassblown cell (FL = 150 mm; red curve).}
\end{figure}

In the experiment, the Rb cell temperature was varied, and the detuning of the 776 nm laser was adjusted to maximize the CBL power while scanning the 780 nm laser. Fig. \ref{fig:fig3}(a) shows the measured CBL spectra at different temperatures for the micromachined cell. For clarity, each lineshape is vertically offset while preserving the relative amplitudes. A reference absorption spectrum of natural Rb is shown in gray.\par

Our results show that the positions of the CBL maxima shift to larger detunings as the temperature increases, consistent with increased Rb vapor density and resonance broadening. As the temperature is raised, the generated CBL power initially increases, reaches an optimum, and then decreases at higher temperatures. This behavior is consistent with increasing reabsorption, which reduces the number of photons that traverse the cell and thus limits the overall efficiency, as well as with possible degradation of phase matching at high vapor density, including Kerr-lensing-induced beam distortion.\par

Fig. \ref{fig:fig3}(b) shows the measured CBL power as a function of the effective gain {$N \cdot L$} for both the micromachined chip-scale cell and the glassblown cell, taken at the constant 780 nm and 776 nm pump powers. For the 150-mm lens, measurements were performed on both cells, whereas the 15-mm lens was used exclusively with the micromachined cell to maximize the CBL output. \par
Notably, employing a 15-mm focal-length lens in a 7-cm cell is impractical because the laser would be significantly defocused and unable to propagate efficiently through the cell. In contrast, a 1.4-mm cell permits stronger focusing of the beam, leading to more efficient CBL generation. \par
Under the optimized conditions we obtained 17 {$\mu$}W of CBL in a micromachined Rb cell using a 15-mm focal-length lens. The chip-scale vapor cell produces higher CBL power than the glassblown cell across the measured {$N \cdot L$} range, indicating a clear performance advantage under our experimental conditions. Importantly, this trend is already observed in the common 150-mm focusing configuration, showing that it does not rely solely on the tighter 15-mm focusing geometry. Several factors likely contribute to this behavior, including the higher operating temperature of the chip-scale cell, stronger confinement of the interaction region, reduced accumulation of angular mismatch over the shorter propagation length, and relaxed alignment constraints within the Rayleigh range.  \par
Measurements performed with the micromachined vapor cell show that the CBL power reaches a maximum at a specific temperature, beyond which the 420 nm emission decreases. This behavior cannot be accounted for by the density-length product alone, since a similar {$N \cdot L$} in
the glassblown cell does not produce a similar decay in emission. This indicates that the elevated local vapor density in the chip-scale cell, arising from its higher operating temperature, modifies the nonlinear interaction. Consequently, the observed saturation is governed not only by the global density-length product but also by local vapor conditions. Likely contributing mechanisms include reabsorption of the generated 420 nm light and degradation of phase-matching in dense vapor at elevated temperatures, both of which impose a limit on the overall conversion efficiency.\par
A comparison of measurements performed with different focusing conditions further shows that the onset of the decay shifts to higher temperatures at increased laser intensity.


For each temperature and corresponding measured CBL power, we independently optimized and recorded the detuning of each pump laser. The zero point for the 776 nm laser was defined at 775.98 nm, while for the 780 nm laser the zero point was set by the \textsuperscript{85}Rb F = 3 absorption line. Fig. \ref{fig:fig3}(c), \ref{fig:fig3}(d), and \ref{fig:fig3}(e) illustrate the optimal detuning of the two laser sources as well as the sum of the detunings. The data are shown for three configurations: a micromachined cell with a 15-mm focal-length lens (blue curve), a micromachined cell with a 150-mm focal-length lens (green curve), and a traditional glassblown cell (red curve).\par
From the measurements of the 780 nm laser offset, we observe that increasing temperature shifts the optimal detuning and modifies the phase-matching condition. We attribute this primarily to temperature-induced broadening of the Rb absorption resonances. The 776 nm laser offset was adjusted accordingly. \par
From the sum of the detunings, we observe a shift of approximately 0.3 GHz between the glassblown and micromachined cell measurements. This shift is likely associated with the higher operating temperature of the micromachined cell, which increases Doppler broadening and can move the condition for the most efficient CBL generation farther from the \textsuperscript{85}Rb F = 3 absorption line. The additional shift of approximately 0.1 GHz observed in the micromachined cell measurements performed with different lenses is more likely caused by the change in laser intensity, which alters the local excitation conditions and therefore the detuning corresponding to maximum CBL generation.\par
Interestingly, near a density-length product of {$1.5\cdot10^{13} 1/cm^2$}, a change in slope is observed. For the micromachined vapor cell, this occurs for both the 780 nm and 776 nm detunings, whereas for the glassblown cell it occurs only for the 776 nm detuning. \par
A plausible interpretation is that the optimal 780 nm detuning is primarily determined by the local response of the strongly absorbing ground-state transition and therefore scales mainly with atomic density N, whereas the optimal 776 nm detuning is more sensitive to the cumulative buildup of the intermediate-state population and the nonlinear interaction along the propagation direction, making it depend more strongly on the density-length product.

\begin{figure}[H]
\centering\includegraphics[width=\textwidth]{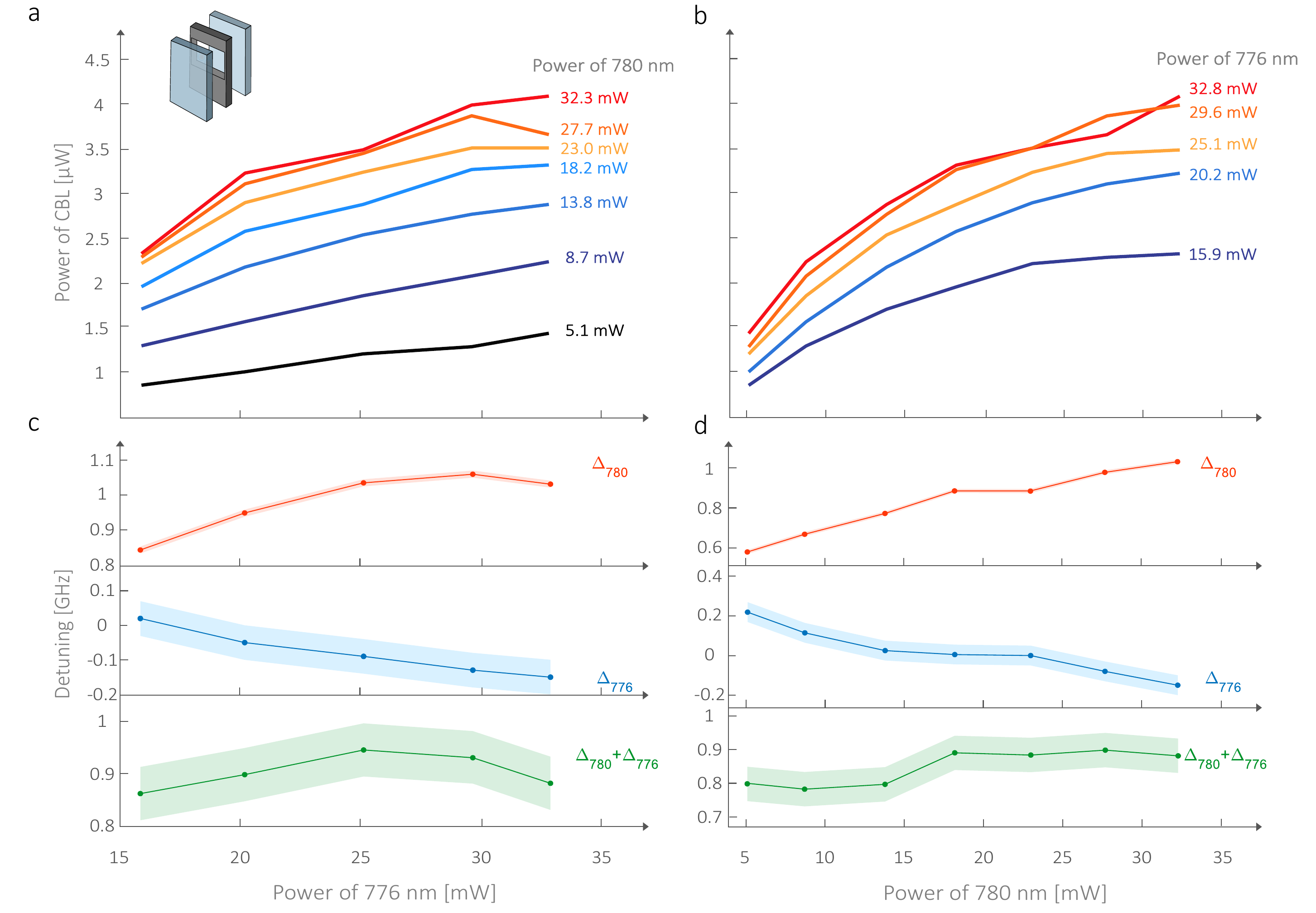}
\caption{\label{fig:fig4} Power dependence of CBL. (a) CBL output vs 776 nm pump power for fixed 780 nm powers (color‐coded). (b) CBL output versus 780 nm power for fixed 776 nm levels (color‐coded). (c, d) Optimal detuning of the 780 nm laser (red), the 776 nm laser (blue), and their sum (green), plotted against (c) power of 776 nm laser at maximum power of 780 nm and (d) power of 780 nm laser at maximum power of 776 nm.}
\end{figure}
Another important scaling parameter is the pump power of the two excitation lasers. To investigate this dependence, we used the 150-mm focusing configuration and alternately varied the powers of the 780~nm and 776~nm lasers and changed their detunings to maximize the CBL power.

Fig.~\ref{fig:fig4}(a) shows the CBL output as a function of the 776~nm power for several fixed values of the 780~nm power. Increasing the 776~nm power leads to a corresponding increase in CBL output, while higher 780~nm power shifts the curves to larger output levels. The 776~nm power is shown starting from 15~mW because lower powers produce insufficient CBL for reliable measurement. The gradual flattening of the curves indicates the onset of saturation, consistent with previous observations \cite{32}.

A similar behavior is observed when varying the 780~nm power. As shown in Fig. \ref{fig:fig4}(b), the CBL output initially increases with 780~nm power but gradually approaches a plateau at higher powers, again indicating saturation of the nonlinear process.

Figures \ref{fig:fig4}(c) and \ref{fig:fig4}(d) show the detuning shifts of the 780~nm and 776~nm lasers, respectively, when operated at maximum power. These measurements indicate that increasing pump power shifts the optimal phase-matching condition, consistent with AC Stark shifts and power-dependent changes in absorption. In a linear approximation, the detuning shifts correspond to a coefficient of approximately 0.01~GHz/mW for both lasers.

These measurements demonstrate that efficient CBL generation requires simultaneous optimization of both pump powers and laser detunings. Increasing the power of one beam alone does not maximize the output unless the other beam and the phase-matching condition is adjusted accordingly.

An important question for FWM-based sources is whether the spectral properties of the pump lasers are transferred to the generated field through the nonlinear medium. This question is particularly relevant in dense, chip-scale vapor cells, where strong nonlinear interaction occurs over a short propagation length and the role of atomic coherence and pump-laser noise in determining the generated linewidth remains an open issue. 

To address this, we measured the spectral linewidth of the generated CBL using a Fabry-Perot cavity with a resolution of 1~MHz.  The transmission profile of the fundamental cavity mode yields a linewidth of $1.4\pm0.2$~MHz (Fig.~\ref{fig:fig1}(c)). Because the analyzer resolution is comparable to the measured width, this value should be interpreted as an upper bound on the intrinsic CBL linewidth. In addition, the 780~nm and 776~nm pump lasers were not frequency stabilized during the measurement, and slow frequency drift during the acquisition can further broaden the observed resonance. The measured linewidth is therefore dominated by instrumental resolution and pump-laser frequency noise. The obtained value is comparable to previously reported CBL linewidths measured in centimeter-scale Rb vapor cells \cite{47}.

\subsection{Coherent mid-IR emission}

The mid-IR emission at 5.2~$\mu$m remains relatively unexplored, largely due to the challenges associated with its detection. To date, only a few studies have reported measurements of CMIRL \cite{17,18,19,20}. In this work we investigate the temperature scaling of the 5.2~$\mu$m emission power in a Si-Pyrex chip-scale Rb vapor cell and compare the measurements with theoretical predictions.

Because the mid-IR signal is weak, due to absorption and scattering in Si, and is not visible, precise adjustment of the cell position, laser detunings, and other parameters is challenging. Instead of optimizing the signal at each temperature, we adopted a different strategy than that used for the CBL measurements. We first optimized the detuning of the 776~nm laser to maximize the CMIRL power at the lowest temperature (solid curve in Fig.~\ref{fig:fig1}(d)), and then increased the cell temperature while keeping this detuning fixed. As a result, the phase-matching condition is not re-optimized at higher temperatures, which partially explains the deviation of the measured data from the ideal scaling.

To estimate the expected CMIRL power, we assume that the photon-pair generation efficiency follows a $\sinh^2(\gamma L)$ dependence. In the ideal case, the amplitude of this function for the CBL is expected to be 12 times larger than that for the CMIRL because of the difference in photon energy. Additional attenuation arises from absorption and scattering in Si and from experimental detection losses. Therefore we estimate a total loss factor of 55 (Fig.~\ref{fig:fig1}(e)), corresponding to a detection efficiency of about 1.8\% of the generated CMIRL power.

Using these parameters, we plot the predicted scaling of the CMIRL power as the dashed curve in Fig.~\ref{fig:fig1}(d). The first three data points lie close to this intrinsic dependence, indicating reasonable agreement with the model, while the remaining points deviate due to the lack of phase-matching re-optimization.

To further investigate the origin of the loss factor, we fabricated a Si wafer with the same cavity geometry and etching procedure used for the Rb vapor cell but without the Pyrex bonding. Using an external 5.2~$\mu$m laser, we measured a transmission of approximately 20\% through the Si structure. Additional losses arise from the difficulty of aligning the 5.2~$\mu$m radiation onto the detector, whose active area is only $1 \times 1$~mm$^2$, as well as from the signal averaging required during the measurements.
  
Notably, we did not observe saturation of the CMIRL power at higher temperatures. We attribute this behavior to the fact that the 5.2 ~$\mu$m transition can benefit from population inversion between the 5D and 6P states, which supports amplification of the generated mid-IR field. As the vapor density increases, this amplification can continue to strengthen the CMIRL output, in contrast to the CBL signal, which is more strongly limited by competing loss mechanisms.


\section{Discussion}
In this work, we demonstrated CW FWM in micromachined chip-scale Rb vapor cells, including the generation of 17 ~$\mu$W of coherent blue emission with a measured linewidth of 1.4 MHz, as well as direct detection of coherent mid-IR emission at 5.2 ~$\mu$m. Comparison with a conventional glassblown vapor cell shows that the chip-scale platform can achieve higher CBL generation efficiency at a comparable density-length product, despite its substantially shorter interaction length. \par

Narrowband 420 nm light is relevant for a range of applications in atomic physics and beyond. For example, an Rb-based frequency reference at 420 nm has been shown to reach a stability of 10\textsuperscript{-15} at 80 s \cite{37}, and both hot- \cite{27,28} and cold-atom \cite{50} Rydberg excitation schemes can make use of 420 nm light. Here, we show that such emission can be generated in a micromachined chip-scale Rb vapor cell while maintaining MHz-level spectral linewidth. This combination of compactness and spectral performance is particularly attractive for atomic and quantum technologies. Injection-locking techniques at the generated wavelength provide a promising route toward further power scaling \cite{49,50}. The strongly reduced footprint of the present source therefore opens a path toward compact photonic and atomic devices based on coherent 420 nm light.\par

The chip-scale platform also enabled direct detection of coherent mid-IR emission at 5.2 $\mu$m and characterization of its temperature scaling. The measured CMIRL power is in good agreement with a simple scaling model once the photon-energy difference and the detection losses are taken into account. The detected mid-IR power was lower than expected, mainly because of transmission losses in the Si-based cell, which we attribute primarily to DRIE-induced surface roughness. Future implementation of direct Si-Si bonding is therefore expected to improve transmission and overall mid-IR performance. Although the linewidth of the 5.2 $\mu$m field was not measured directly, the blue and mid-IR fields are generated in the same phase-matched FWM process and are linked by energy conservation, so their frequency fluctuations are expected to be strongly correlated.\par

From an application perspective, the generated CMIRL represents an interesting alternative to conventional mid-IR sources such as quantum cascade lasers (QCLs). While QCLs offer high power and broad tunability, their free-running frequency is governed by current and temperature and typically requires external stabilization. By contrast, the CMIRL demonstrated here is generated on an atomic transition in Rb and is therefore naturally linked to an atomic resonance. This makes it a promising compact narrowband mid-IR source. Its main limitation is the lack of broad continuous tunability, since the emission is fixed by the discrete Rb transition near 5.2 $\mu$m. Accordingly, this source is especially well suited for use as a compact frequency reference or as a seed for higher-power mid-IR systems.\par

Overall, these results establish micromachined Rb vapor cells as a promising platform for compact nonlinear frequency conversion. In addition to supporting efficient generation of both visible and mid-IR light, the platform is compatible with standard microfabrication and with different window materials, enabling flexible device design for multi-wavelength operation.  Wafer-based microfabrication of the cells enables scalable and reproducible fabrication, in contrast to conventional individually assembled vapor cells, and provides a direct pathway for integration with photonic integrated circuits. This combination of nonlinear functionality, compactness, and scalability supports integration into atomic sensors, clocks, and related chip-scale photonic devices.

\section{Back matter}

\begin{backmatter}

\bmsection{Acknowledgment}
The authors thank Roy Zektzer for fruitful discussions. They also thank Arieh Grosman and Kfir Levi for their assistance with cell fabrication.

\bmsection{Disclosures}
The authors declare no conflicts of interest.

\bmsection{Data availability} Data underlying the results presented in this paper are not publicly available at this time but may be obtained from the authors upon reasonable request.

\end{backmatter}

\bibliography{sample}

\end{document}